# MLatom software ecosystem for surface hopping dynamics in Python with quantum mechanical and machine learning methods

Lina Zhang,[1] Sebastian V. Pios,[2†] Mikołaj Martyka,[3†] Fuchun Ge,[1†] Yi-Fan Hou,[1†] Yuxinxin Chen,[1] Lipeng Chen,[2] Joanna Jankowska,[3*] Mario Barbatti,[4,5*] Pavlo O. Dral[1,6,7,8*]

[1]*College of Chemistry and Chemical Engineering, Xiamen University, Xiamen, Fujian 361005, China*

[2]*Zhejiang Laboratory, Hangzhou, Zhejiang 311100, People's Republic of China*

[3]*Faculty of Chemistry, University of Warsaw, Pasteura 1, Warsaw, 02-093, Poland*

[4]*Aix Marseille University, CNRS, ICR, 13397 Marseille, France*

[5]*Institut Universitaire de France, 75231, Paris, France*

[6]*State Key Laboratory of Physical Chemistry of Solid Surfaces, Xiamen University, Xiamen, Fujian 361005, China*

[7]*Innovation Laboratory for Sciences and Technologies of Energy Materials of Fujian Province (IKKEM), Xiamen University, Xiamen, Fujian 361005, China*

[8]*Fujian Provincial Key Laboratory of Theoretical and Computational Chemistry, Xiamen, Fujian 361005, China*

Emails: joanna.jankowska@uw.edu.pl, mario.barbatti@univ-amu.fr, dral@xmu.edu.cn

[†]*Equal contribution*

**Abstract**

We present an open-source MLatom@XACS software ecosystem for on-the-fly surface hopping nonadiabatic dynamics based on the Landau–Zener–Belyaev–Lebedev (LZBL) algorithm. The dynamics can be performed via Python API with a wide range of quantum mechanical (QM) and machine learning (ML) methods, including *ab initio* QM (CASSCF and ADC(2)), semi-empirical QM methods (*e.g.*, AM1, PM3, OMx, and ODMx), and many types of machine learning potentials (*e.g.*, KREG, ANI, and MACE). Combinations of QM and ML methods can also be used. While the user can build their own combinations, we provide AIQM1, which is based on Δ-learning and can be used out of the box. We showcase how AIQM1 reproduces the isomerization quantum yield of *trans*-azobenzene at a low cost. We provide example scripts that, in a dozen lines, enable the user to obtain the final population plots by simply providing the initial geometry of a molecule. Thus, those scripts perform geometry optimization, normal mode calculations, initial condition sampling, parallel trajectories propagation, population analysis, and final result plotting. Given the capabilities of MLatom to be used for training different ML models, this ecosystem can be seamlessly integrated into the protocols building ML models for nonadiabatic dynamics. In the future, a deeper and more efficient integration of MLatom with Newton-X will enable vast range of functionalities for surface hopping dynamics, such as fewest-switches surface hopping, to facilitate similar workflows via the Python API.

**Introduction**

Nonadiabatic dynamics simulations are essential for understanding the photophysical and photochemical processes.[1-8] Among various types of such simulations, on-the-fly surface hopping dynamics is the most widely used.[1, 3, 6-8] In this simulation type, a swarm of independent trajectories is propagated, each using the forces for a single adiabatic electronic state. The nonadiabatic nature of the dynamics is recovered by allowing the trajectories to hop with some probability to another electronic state surface. Different algorithms that vary in many aspects, particularly how the hopping probabilities are calculated, were suggested.

Among these many algorithms, the Landau–Zener approximation[9-11] based on the Belyaev–Lebedev formulation in the adiabatic representation[12] is gaining momentum because it avoids explicit calculation of the nonadiabatic couplings, hence speeding up the simulations while also enabling surface hopping even for methods where these couplings are not available.[13-16] We call this algorithm LZBL surface hopping in the following. Hopping probabilities in LZBL surface hopping are obtained from the potential energy surface topography, and, in this sense, it is similar to Baeck–An,[17] κTSH,[18] and Zhu–Nakamura[19] surface hopping, which are also wave-function free, topography-based algorithms. The LZBL approximation is easy to implement, and the dynamics can be propagated to an arbitrary number of electronic states. Its limitation is that LZBL is only reliable if a maximum of two adiabatic states are close in energy.[13] However, despite this known limitation, LZBL dynamics has been shown to agree well with less approximate fewest-switches surface hopping dynamics for many systems of practical interest.[13]

The usefulness of the LZBL approximation led to its implementations in several surface-hopping packages such as ZagHop,[20] ABIN,[13] PySurf,[14] and Libra.[21] These packages have interfaces to the third-party electronic structure software needed to calculate the energies and forces for the electronic states involved with the quantum mechanical (QM) methods. The progress in machine learning (ML), particularly in the context of surface-hopping dynamics[15, 16, 22-43] (see also reviews[44-49]), shows the potential of substituting slow QM with fast ML models for evaluating forces and energies. This potential is underutilized for the LZBL approximation, although there is a growing interest in using ML to accelerate LZBL surface hopping dynamics. So far, its utility has been investigated for surface hopping simulations of the phenanthrene molecule for one type of potentials (SchNet[50]) trained on the semi-empirical QM data provided by TD-DFTB.[15] Independently, the data-oriented package PySurf was developed, and its utility was demonstrated in LZBL surface hopping for three and five-dimensional systems. The corresponding article[14] reported a simple interpolation scheme and alluded to the prospects of using ML. Some of us also reported an ad-hoc modification of the ZagHop package for ML-accelerated LZBL dynamics.[16]

It is apparent, though, that the rapidly developing field of ML methods applied to surface hopping needs a versatile software ecosystem that would enable easy use, modification, and extension. Such a versatile ecosystem, based on the MLatom package,[51] exists for a wide variety of computational chemistry applications but not for nonadiabatic dynamics. The previously reported Newton-X and MLatom ecosystem was limited to command-line input and one type of ML potentials, non-LZBL approximations, while also having a relatively inefficient disk I/O communication.[22, 41] Software packages such as SchNarc,[34] PyRAI(2)MD,[30] Jade,[40] and others used in the original studies,[15, 52] often specialize on a specific type of ML potential and are limited to surface-hopping dynamics.

Here, we report one more step towards a versatile software ecosystem based on MLatom[51, 53] by offering a Python API for users to efficiently build surface-hopping workflows, from initial geometry to initial conditions sampling (also with routines adapted from Newton-X[22]) to dynamics and population analysis (Figure 1). The ecosystem currently supports LZBL surface hopping (LZSH for short) dynamics with a wide variety of ML models and with *ab initio* and reduced-cost semi-empirical QM methods. In addition, it provides unique access to performing surface-hopping dynamics with the general-purpose AIQM1 method, which is based on the semi-empirical ODM2 method[54] with a Δ-learning[55] ML correction.[56] AIQM1 was so far only used for calculating vertical excitation energies[56] and optimizing excited-state geometries[56, 57] (apart from the known successful applications of AIQM1 in ground-state simulations[56-58] including dynamics[59]). The user can build their custom models with different combinations of ML and QM methods. This versatility is based on the coherent data formats we developed for MLatom.

In addition to the standard algorithm for LZBL dynamics, our implementation supports the modification where the probability of the back hopping is calculated with the reduced kinetic energy reservoir, as it was shown to be important when nonadiabatic coupling vectors are not provided by the electronic structure calculations.[60]

In the sections below, we provide the details and showcase the applications, including the short, all-in-one Python script showing how the ecosystem can be used, from setting up the simulations to obtaining the production result.

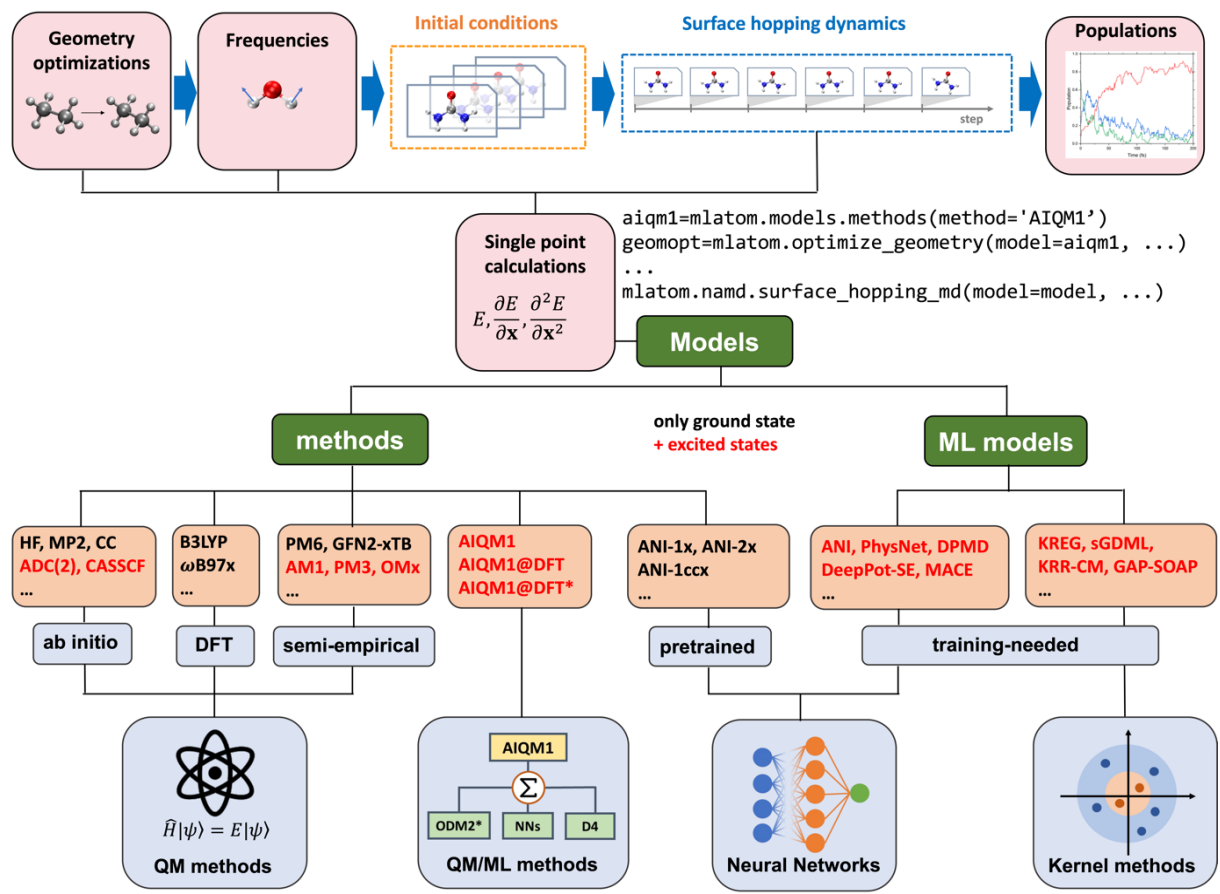

**Figure 1.** The MLatom software ecosystem provides seamless workflows for performing the surface hopping dynamics from setting initial conditions (geometry optimizations, normal mode analysis, and phase-space sampling) to propagating trajectories and analyzing them, culminating with the plotting of populations. These steps can be performed with appropriate QM methods, ML models, and hybrid ML/QM techniques. Models supporting excited-state calculations are shown in red. Initial conditions can also be obtained with routines adapted from Newton-X. Some elements of this figure are adapted from Ref. 51 with permission. Copyright 2024, the Authors.

**Methods**

*Implementation of the data format for electronic states*

The key to the seamless integration of the disparate parts is the unified communication between the parts through the standard data format. We have extended MLatom's format[51] to incorporate the information from the excited-state calculations in an intuitive and easily accessible way. This data format is based on Python classes dedicated to molecule objects, molecular databases collecting the molecule objects, and molecular dynamics trajectories. These objects can be dumped and loaded in different formats, from text files to JSON and H5MD[61] files.

The cornerstone of the data format is the molecule object, which contains information about an atomistic system and is not limited to a single molecule. Ground-state properties such as energies are directly accessible in the molecule objects, e.g., in Python as `molecule.energy`. The challenge with the excited states is that many such properties correspond to different electronic states (e.g., energies for $S_0$ and $S_1$ states) and transitions between them (e.g., oscillator strengths, excitation energies, or energy gaps). In addition, different electronic states might have different spin states, e.g., $S_1$ and $T_1$. All this information must be encoded and easily accessible in the molecule object.

To incorporate all these requirements, we introduce the convention that each electronic state is considered a separate molecule object containing all the available properties associated with this state (spin state, energies, gradients, *etc*.). These molecule objects for each electronic state are then appended to the main molecule object. They can be accessed in Python as `molecule.electronic_states`. The transition properties (oscillator strengths, excitation energies from the ground state, and gaps between states) are saved in the main molecule object with the order corresponding to that in `molecule.electronic_states`. The structure of the resulting molecule object and an example in Python are shown in Figure 2.

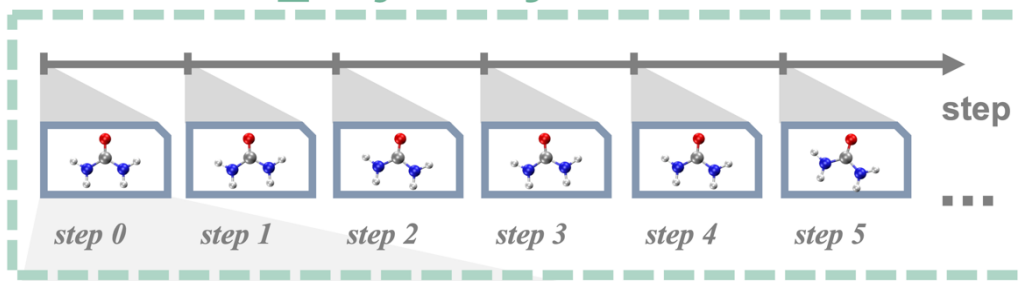

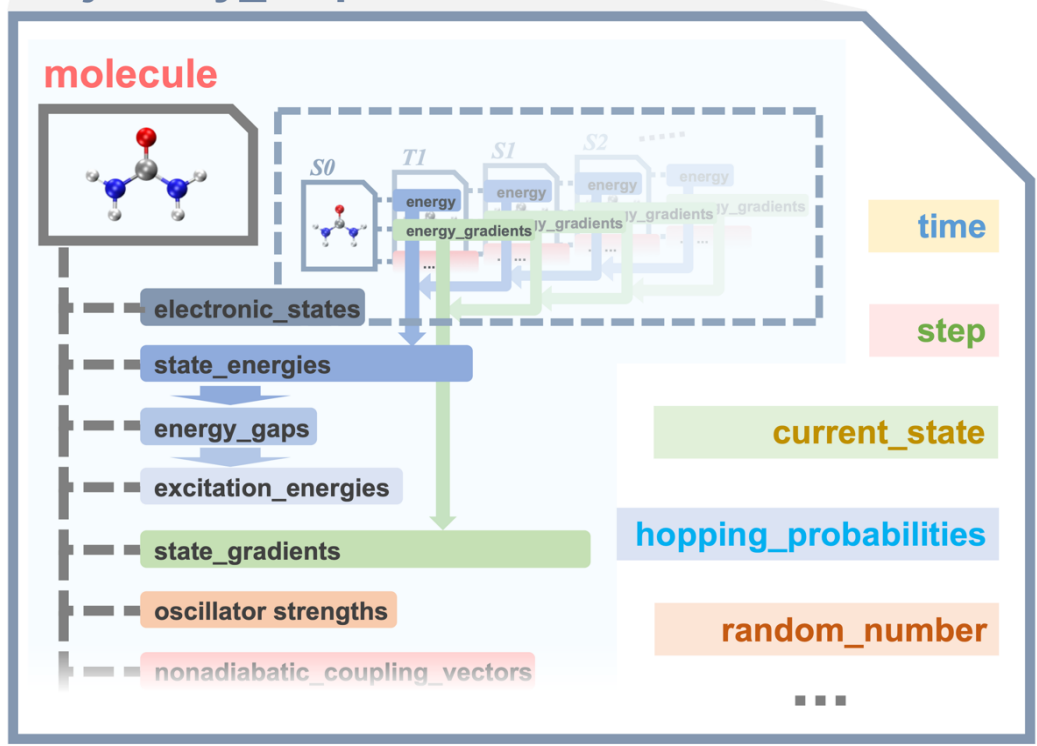

**Figure 2.** Data format for excited-state properties and code snippet for accessing them in Python.

Molecular dynamics trajectory objects were extended to incorporate information related to the surface hopping dynamics for each time step, i.e., the current electronic state, hopping probability, and nonadiabatic couplings (for future extensions) (Figure 2). Each time step of the trajectory objects contains the molecule object with the other relevant information described above.

*Implementation of single-point excited-state calculations*

Surface hopping dynamics requires a single-point evaluation of energies and energy gradients (and nonadiabatic coupling vectors for algorithms not discussed here) for a fixed nuclear geometry at each time step. These properties are commonly calculated with the appropriate QM methods for the required electronic states (energy gradients are often calculated only for the current state). An increasing number of studies use ML models to speed up the simulations by completely replacing or invoking QM methods only for the critical regions.[15, 16, 22-49] The general-purpose AIQM1 method[56] was also used for a related task for excited-state geometry optimizations,[56, 57] but this only requires energy gradients and energies for one state throughout the calculations. At the same time, surface-hopping dynamics may switch between states. One of the biggest practical challenges is that these QM and ML methods are implemented in different software packages without standard, unified formats.

This rapidly increasing diversity of the possible but incompatible solutions for single-point calculations necessitates the development of a standard way to call routines for generating the required properties. The code should also be easily extendable for fast prototyping and development. We developed a highly flexible standard accessible via Mlatom's Python API, which allows for building arbitrarily complex routines that populate the molecule objects with the properties specified in standard arguments. Many interfaces require special arguments, too, which are accounted for in the implementation. Examples of the code using this Python API for single-point excited-state calculations with several QM and ML approaches are given in Figure 3.

```
Define QM methods

# define AIQM1/CIS
model = ml.models.methods(method='AIQM1')

# define ODM2/CIS
model = ml.models.methods(method='ODM2')

# define ODM2/MRCI where more options must be provided
model = ml.models.methods(method='ODM2', program='mndo',
                          read_keywords_from_file = 'mndokw')

# define AIQM1/MRCI where more options must be provided
model = ml.models.methods(method='AIQM1', qm_program='mndo',
                          qm_program_kwargs={'read_keywords_from_file': 'mndokw'})

# define CASSCF where more options must be provided
model = ml.models.methods(program='columbus',
           command_line_arguments=['-m', '1700'],
           directory_with_input_files='columbus'))

# define ADC(2) where more options must be provided
model = ml.models.methods(program='turbomole',
           directory_with_input_files='turbomole'))
```

```
Define ML or another custom model, e.g., ML/QM

# To define a custom model, the user must build a class
class mlmodels():
    def __init__(self, nstates = 5):
        self.models = [None for istate in range(nstates)]
        for i in range(nstates):
            self.models[i] = [ml.models.ani(model_file=f'ani{i}.pt')]

    def predict(self,molecule=None,nstates=5,current_state=0,
        calculate_energy=True,calculate_energy_gradients=True):

        molecule.electronic_states = [molecule.copy() for i in range(nstates)]
        for istate in range(nstates):
            moltmp = molecule.electronic_states[istate]
            self.models[istate].predict(molecule=moltmp,
                calculate_energy = True, calculate_energy_gradients = True)

        molecule.energy = molecule.electronic_states[current_state].energy
        molecule.energy_gradients = molecule.electronic_states[
                                    current_state].energy_gradients

model = mlmodels()
```

```
Calculate electronic state properties with the given model

model.predict(molecule=mol, nstates=3,
              current_state=2, # the state numbering starts from 0
              calculate_energy=True,
              calculate_energy_gradients=True)
              # alternatively, define for which states to calculate:
              # calculate_energy_gradients=[True, True, True])
```

**Figure 3.** Single-point calculations using the standardized Mlatom Python API for several QM and ML approaches.

Mlatom currently supports several quantum mechanical approaches, ADC(2)[62, 63] via the interface to Turbomole,[64] CASSCF via the interface to Columbus,[65, 66] and a range of semi-empirical QM methods with different Hamiltonians (AM1,[67] PM3,[68] OMx,[54] ODMx,[54] *etc*.) and configuration interaction (CI) types (variants of spin-flip CIS[69, 70] and GUGA-CI[71] referred to as MRCI in literature) provided by the MNDO program[72] (Figure 1). In addition, we implemented AIQM1 with different CI types by interfacing MLatom to the MNDO, providing the ODM2* semi-empirical part (ODM2 method without dispersion correction which is replaced with a newer type of dispersion correction in AIQM1),[54, 56] TorchANI[73] for the neural network (NN) corrections of the modified ANI-type,[73] and dftd4[74] for D4-corrections.[75] We note that all the constituent terms of AIQM1 are also stored in the molecule object. Only the semi-empirical QM part (ODM2*) gives different predictions for different electronic states; other corrections are the same for all states. While the ground state's gradients and energies at AIQM1 are approaching the coupled-cluster level, the vertical excitation energies are of semi-empirical QM quality. The energy gradients are corrected for each state with the same NN and D4 corrections, leading to different dynamics at AIQM1 than those of the underlying ODM2* method.

A variety of ML models are supported as shown in Figure 1: the (p)KREG,[76, 77] MACE,[78, 79] ANI,[73] DPMD[80] and DeepPot-SE[81], GAP[82]-SOAP[83], sGDML,[84] and PhysNet[85] via the interfaces to the corresponding programs as described elsewhere.[51] Currently, we have not implemented specific models for excited states. Instead, we built a new model from individual ML models for energies and gradients at each electronic state (and oscillator strengths if required); see Figure 3 for an example.

*Initial conditions generation*

To start the trajectory, we need initial conditions. They can be sampled randomly from the Maxwell–Boltzmann distribution or from the ground-state trajectory, which can be efficiently run using many different QM and ML methods with MLatom.[51] A widespread (and, de facto, default) approach is to generate initial conditions via sampling a harmonic oscillator Wigner distribution, which accounts for the zero-point energy.[86, 87] This is done in Newton-X with the `initcond` routine,[22] but the output is in the text format and requires post-processing scripts to import the initial conditions for the calculations in Python. Hence, we made Newton-X's Wigner sampling accessible via MLatom's Python API by rewriting the code in Python.

Our implementation also supports the selection of initial conditions for the given window in the absorption spectrum from the Wigner sampling analogously to Newton-X.[22] This selection requires the calculation of excitation energies and oscillator strengths, which must be provided by MLatom's model, as shown for single-point calculations.

In addition, harmonic Wigner sampling can be particularly problematic for low-frequency normal modes,[88] whose anharmonicity may lead to unphysical structures, such as overstretched C–H bonds. In this case, we suggest setting all low frequencies to 100 cm$^{-1}$ if they are below this value. A similar procedure is commonly applied when calculating thermochemical properties.[89]

The overview of the initial conditions sampling strategies and the code snippets are given in Figure 4.

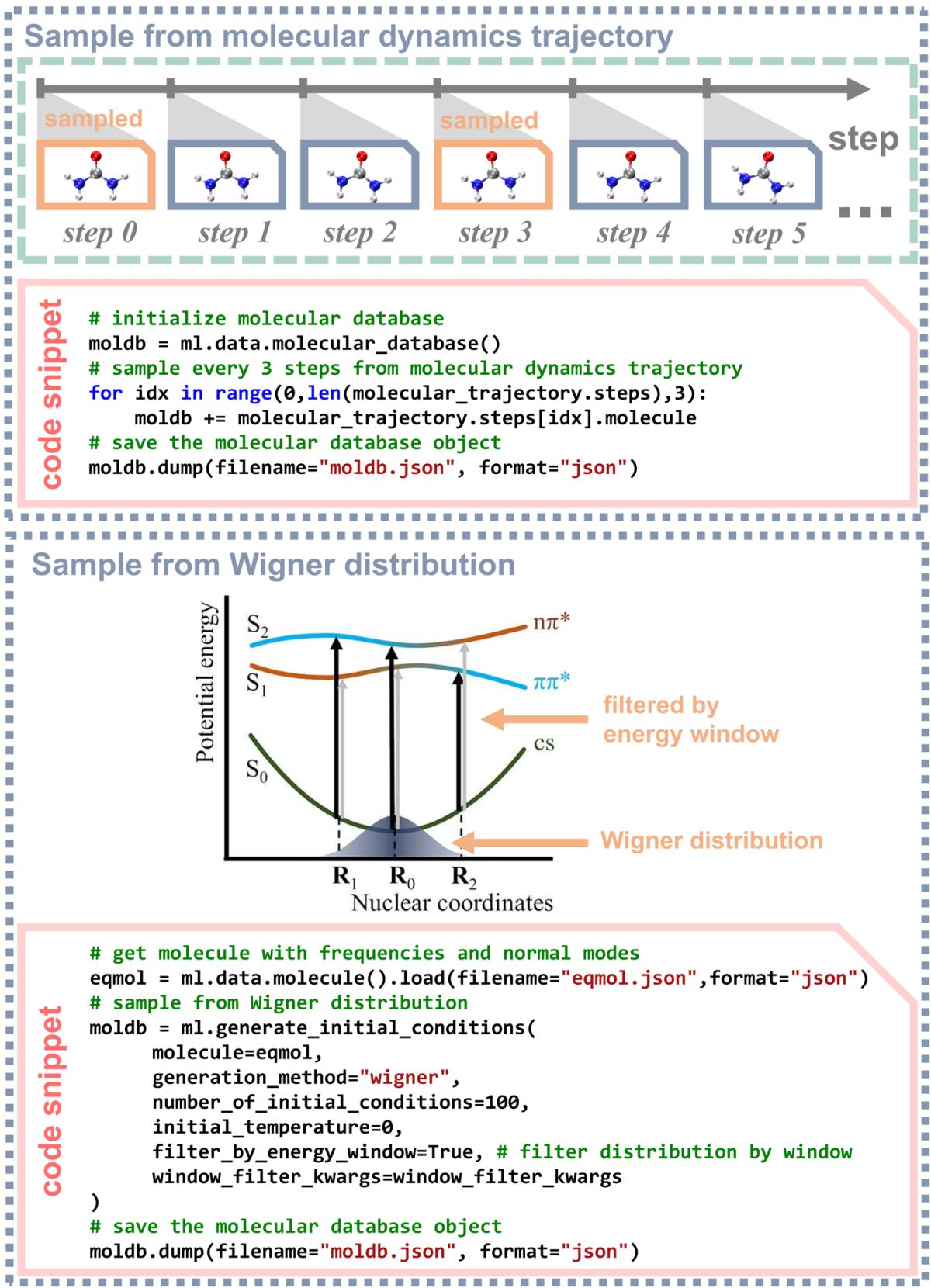

**Figure 4.** Generation of initial conditions with MLatom via sampling from the ground-state dynamics and the Wigner distribution with and without filtering by the excitation energy window. The panel 'Sample from Wigner distribution' reproduced from Ref. [22] with permission. Copyright 2022, the Authors.

One more convenient feature of our implementation is that the normal mode sampling requires their calculations preceded by geometry optimizations. All of these can be done with MLatom's existing Python API at various QM methods and ML models, including AIQM1, which is faster and more accurate than, e.g., common DFT for such simulations.[56-59] In the normal mode sampling, we are only interested in the vibrational modes, not rotational and translational ones. Hence, we need to remove them. For this, we use the interface with Gaussian[90] and PySCF[91-93], which remove rotational and translational modes.

*Landau–Zener–Belyaev–Lebedev surface hopping dynamics*

LZBL surface hopping is one of the most popular ways to avoid the computation of the nonadiabatic couplings when evaluating hopping probability. In the original Landau–Zener algorithm,[9-11] the nonadiabatic transition probability $P_{j \to k}^{\text{LZSH}}$ between the adjacent adiabatic states $j$ and $k$ is given by the following formula based on the diabatic representation,

$$P_{j \to k}^{\text{LZSH}} = \exp\left(-\frac{2\pi H_{ab}^2}{\hbar \dot{R} |H_{bb}' - H_{aa}'|}\right), \tag{1}$$

where $H_{ab}$ is the off-diagonal matrix element between diabatic states $a$ and $b$, while $H_{aa}$ and $H_{bb}$ are diagonal matrix elements and $H_{aa}'$ and $H_{bb}'$ are derivatives of $H_{aa}$ and $H_{bb}$ with respect to nuclear coordinate $R$. $\dot{R}$ in the formula denotes the time derivative of $R$. This formula does not require information on nonadiabatic coupling vectors. However, other difficulties arise due to the use of the diabatic representation. In 2011, Belyaev and Lebedev[12] rewrote the Landau–Zener formula based on the adiabatic representation:

$$P_{j \to k}^{\text{LZSH}} = \exp\left(-\frac{\pi}{2\hbar}\sqrt{\frac{Z_{jk}^3}{\ddot{Z}_{jk}}}\right). \tag{2}$$

In this new formula, $Z_{jk}$ is the absolute energy gap between two adiabatic states and $\ddot{Z}_{jk}$ is the second-order time derivative of $Z_{jk}$. After proposing this LZBL formula, many researchers adopted it to run surface hopping dynamics due to its simplicity.[13-16] In the practical application, $\ddot{Z}_{jk}$ is often approximated by the three-point finite difference formula for the middle point if it is a minimum.[13] The difficulty with this approximation is that it requires information about the previous and future time steps, which hampers its implementation in the existing surface-hopping packages as it might require substantial code rewriting.

In MLatom, we have implemented the LZBL algorithm and also use a three-point finite difference formula to approximate the $\ddot{Z}_{jk}$ term. This implementation is rather straightforward as we can take advantage of the convenience of the modular design of the program and Python API in MLatom, where all the data and routines are handled in standard formats. MLatom's native implementation of molecular dynamics (MD) is also used as the trajectory propagator. It is beneficial because we can reuse the existing routines to write a surface-hopping code without rewriting or modifying it. To evaluate the probability at the current time step, we need to calculate the time derivative of the energy gap by evaluating this property at the previous and future time steps. MLatom's MD propagator propagates the short trajectory at the current surface starting from the current time step and returns the short trajectory object with the future time step. LZBL routine stores the trajectory object with information about the past steps. Suppose the energy gap between the current surface and a specific surface for the current step is the smallest among the three gaps (at the past, current, and future time steps). In that case, the LZBL hopping probability can be evaluated. The surface will be switched when the probability exceeds the random number, and hopping is allowed. This future point is easily integrated into the LZBL trajectory if no hopping happens. If the hopping happens, the future step has to be discarded as the energy gradients for the current step need to be evaluated at another surface. Velocities must be adjusted to ensure the total energy conservation after hopping.

The velocities are adjusted by rescaling them along the momentum direction if there is sufficient kinetic energy to compensate for the change in the potential energy. If there is insufficient energy, the hopping is rejected, which may happen when hops to a higher energy surface are attempted (back hopping). The number of back-hoppings in momentum-rescaled velocities may be too large; hence, recently, it was suggested to reduce the amount of available kinetic energy for the back-hopping by dividing the kinetic energy by the number of degrees of freedom.[60] We implemented this option for the LZBL surface hopping in MLatom, too, and we will use it for some applications later in this article.

Using MLatom's API to run the LZBL surface hopping has another critical advantage in high-performance dynamics propagation with ML models: As they are seamlessly integrated into the Python code, all single-point evaluations can be performed directly in RAM. This is in contrast to many existing platforms, including our own previously reported Newton-X interface to MLatom,[22, 41] where the dynamics driver and packages for single-point calculations communicate via extensive disk I/O because, during the original implementation of older surface-hopping packages, the major bottle-neck was the slow single-point calculations with QM methods. The single-point calculations with the ML models are so fast that the bottleneck may become the communication with the dynamics driver, which is the case for platforms using disk I/O for communication. Hence, MLatom's surface-hopping is not just user-friendly because of the use of Python API; it is also highly efficient. In addition, we can efficiently calculate many (hundreds and thousands) dynamics trajectories in parallel on a single computing node, especially for ML models.

**Applications**

Bringing it all together, Figure 5 shows a typical Python script that uses the MLatom ecosystem for running end-to-end surface-hopping simulations starting from the initial geometry guess and ending with the population plot. This script is easily adjusted by choosing the required combinations of QM methods and ML models and settings for initial conditions sampling and dynamics.

```python
import mlatom as ml

# Load the initial geometry of a molecule
mol = ml.data.molecule()
mol.charge=1
mol.read_from_xyz_file('cnh4+.xyz')

# Define models
aiqm1 = ml.models.methods(method='AIQM1',
                qm_program_kwargs={'save_files_in_current_directory': True,
                                   'read_keywords_from_file': f'mndokw'})
method_optfreq = ml.models.methods(method='B3LYP/Def2SVP', program='pyscf')

# Optimize geometry
geomopt = ml.simulations.optimize_geometry(model=method_optfreq,
                                           initial_molecule=mol)
eqmol = geomopt.optimized_molecule
eqmol.write_file_with_xyz_coordinates('eq.xyz')

# Get frequencies
ml.simulations.freq(model=method_optfreq,
                    molecule=eqmol)
eqmol.dump(filename='eqmol.json', format='json')

# Get initial conditions
init_cond_db = ml.generate_initial_conditions(molecule=eqmol,
                                    generation_method='wigner',
                                    number_of_initial_conditions=16,
                                    initial_temperature=0)
init_cond_db.dump('test.json','json')

# Propagate multiple LZBL surface-hopping trajectories in parallel
# .. setup dynamics calculations
namd_kwargs = {
            'model': aiqm1,
            'time_step': 0.25,
            'maximum_propagation_time': 5,
            'hopping_algorithm': 'LZBL',
            'nstates': 3,
            'initial_state': 2,
            }

# .. run trajectories in parallel
dyns = ml.simulations.run_in_parallel(molecular_database=init_cond_db,
                                    task=ml.namd.surface_hopping_md,
                                    task_kwargs=namd_kwargs,
                                    create_and_keep_temp_directories=True)
trajs = [d.molecular_trajectory for d in dyns]

# Dump the trajectories
itraj=0
for traj in trajs:
    itraj+=1
    traj.dump(filename=f"traj{itraj}.h5",format='h5md')

# Analyze the result of trajectories and make the population plot
ml.namd.analyze_trajs(trajectories=trajs, maximum_propagation_time=5)
ml.namd.plot_population(trajectories=trajs, time_step=0.25,
                    max_propagation_time=5, nstates=3, filename=f'pop.png')
```

**Figure 5.** Python scripts showcasing end-to-end workflow for surface hopping dynamics with MLatom software ecosystem. This example uses routines for initial conditions adapted from Newton-X.

Below, we showcase specific examples of this workflow with different combinations of QM and ML techniques for several molecules: methaniminium cation $CNH_4^+$, pyrazine, trans-azobenzene featuring photoisomerization, and a rather large ferro-wire system with 80 atoms. We show the use of quantum mechanical CASSCF, ADC(2), AIQM1/CIS, ODM2/CIS, AIQM1/MR-CISD, and ODM2/MR-CISD methods as well as ANI machine learning potentials for single-point calculations.

*Methaniminium cation $CNH_4^+$*

As the first application, we show that the platform can perform the surface-hopping dynamics with both CASSCF and AIQM1/CI methods of methaniminium cation $CNH_4^+$. The calculations were done with SA-3-CASSCF(12,8)/6-31G(d) and AIQM1/MR-CISD(4,3) and using the full kinetic energy reservoir (see Computational details). Remarkably, the populations at the AIQM1 level are in rather good agreement with the CASSCF result (Figure 6). However, it uses the semi-empirical Hamiltonian for calculating excitation energies. The speed of the AIQM1 simulations is also much higher than that of CASSCF: AIQM1 calculations took just 4 hours for 100 trajectories propagated in parallel on 36 CPU cores (Intel(R) Xeon(R) Gold 6240 CPU @ 2.60GHz), while CASSCF took 32 hours. Both trajectories used a 0.05 fs time step and propagated for 100 fs. The gains are much larger for bigger systems, making AIQM1 a cost-effective, out-of-the-box method for surface-hopping simulations.

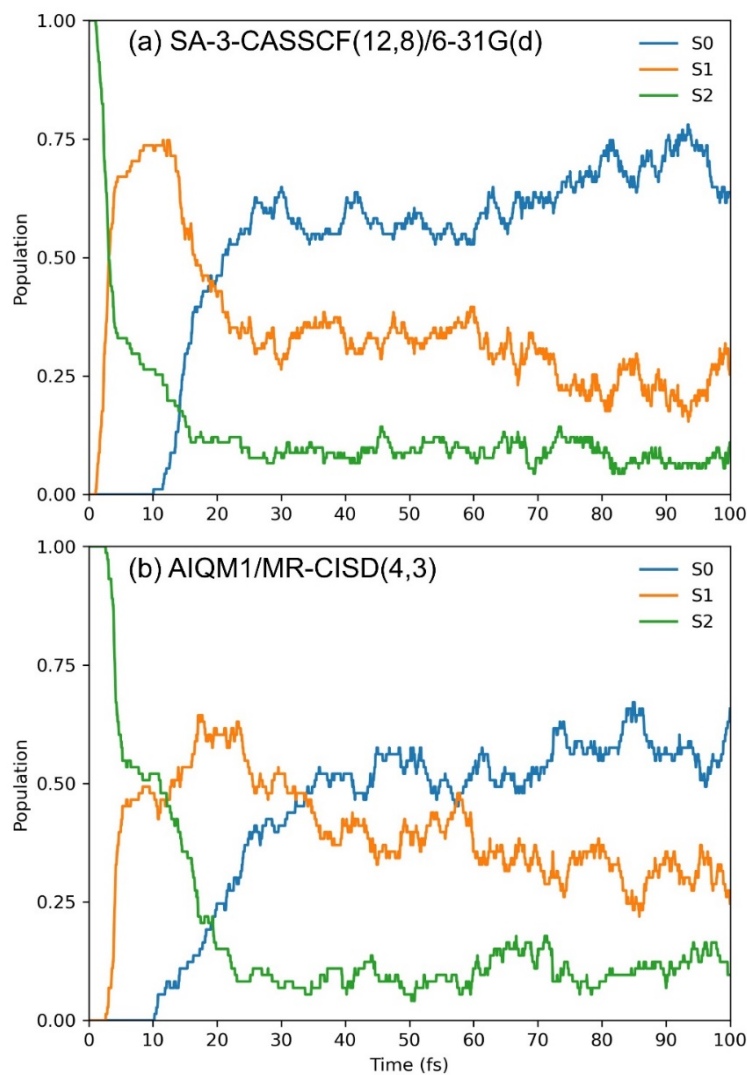

**Figure 6.** The population dynamics with CASSCF (top) and AIQM1/MRCI obtained from the surface-hopping dynamics of methaniminium cation $CNH_4^+$ within the LZBL formalism.

*Pyrazine*

Pyrazine is a popular system for investigating the performance of various QM and ML approaches.[14, 16, 94] It was shown that ADC(2) provides a good description of its dynamics[94] that can also be accelerated with ML[16]. The population dynamics of pyrazine recalculated with our platform and ADC(2)/cc-pVDZ (Figure 7) is very similar to the previously reported dynamics with the larger basis set aug-cc-pVDZ. We also propagate trajectories with a much faster AIQM1/CIS method to show the program's versatility. While the main features are similar, the AIQM1/CIS population is a bit off quantitatively at this approximate level as the population dynamics is overall slower, and the $S_4$ state is overpopulated.

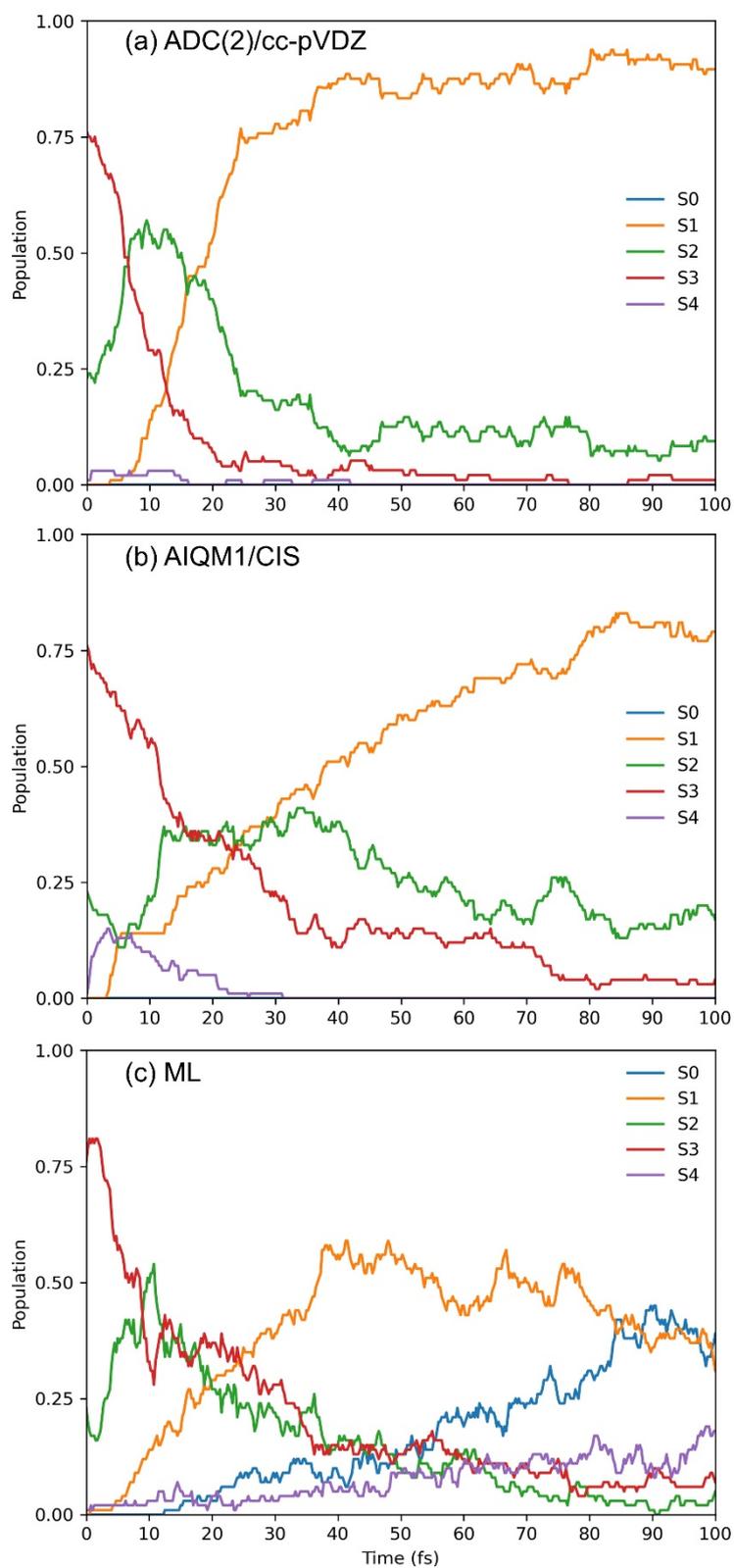

**Figure 7.** The population dynamics with a) ADC(2), b) AIQM1/CIS, and c) ML potentials ANI obtained from the surface-hopping dynamics of pyrazine within the LZBL formalism.

To demonstrate the surface-hopping dynamics with ML, we train individual ANI models for each electronic state by using only the 250 points sampled from the Wigner distribution and the custom models shown in Figure 3. The reference energies and energy gradients were generated at the ADC(2)/cc-pVDZ level. The populations are shown in Figure 7, and they reproduce the reference ADC(2) dynamics quite well at the beginning of the dynamics. Still, they start to deteriorate for a longer propagation time, as expected, given that the models were only trained on the initial conditions. Interestingly, initially, the ML populations are not worse than the AIQM1/CIS populations. We will report active learning procedures yielding better ML models separately; here, we just focus on the implementation of the surface-hopping algorithm.

The computational cost of three different simulations varies from very slow ADC(2) which takes two days on 95 CPUs (Platinum 8253 2.20GHz), AIQM1/CIS with moderate cost (about 2 hours on 36 Intel(R) Xeon(R) Gold 6240 CPU @ 2.60GHz CPUs), and speedy ML propagation taking 3 minutes on 16 CPUs (Intel(R) Xeon(R) Gold 6226R CPU @ 2.90GHz). All trajectories were propagated with the same settings, i.e., we propagated 100 trajectories for 100 fs with 0.25 fs time step and using full kinetic energy reservoir (see also Computational details). Each trajectory was propagated on a single CPU thread.

*Isomerization dynamics of trans-azobenzene*

To showcase the ability of the LZBL surface hopping (LZSH) to simulate photoreactions, we also investigate a classical problem in photochemistry: the photoisomerization of *trans*-azobenzene, after its $n \rightarrow \pi^*$ excitation to S$_1$ (Figure 8). To this end, we have propagated 3 sets of surface hopping trajectories: (a) using LZSH at the ODM2/MR-CISD level, (b) LZSH at AIQM1/MR-CISD, and (c) FSSH at ODM2/MR-CISD. The details of the semi-empirical MRCI calculations can be found in the Computational Details section.

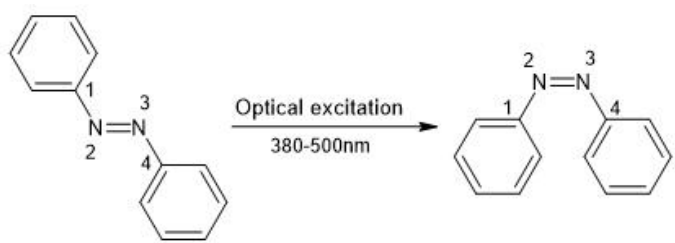

**Figure 8.** Photoisomerization of *trans*-azobenzene.

All sets of trajectories started from the same initial conditions, generated by Wigner sampling, with geometry optimization and vibrational frequencies computed at the AIQM1 level of theory. The sampled geometries were later filtered to an excitation window of 2.53 ± 0.3 eV corresponding to the $S_0 \rightarrow S_1$ transition energy at the AIQM1-optimized ground state minimum. 100 initial conditions were selected for each set and propagated for 1000 fs with a 0.1 fs time step.

The evolution of the $S_0$ state population in each set of trajectories is shown in Figure 9. Quantitatively, all 3 sets show a very similar picture of the ground-state deactivation dynamics. To extract the characteristic timescales for each set of trajectories, an exponential function $f(t)$ was fitted to the ground-state population rise, with $f(t) = A(1 - \exp(-t/\tau))$. where the $\tau$ is the excited-state lifetime and the parameter $A$ of the fit can be associated with the infinite-time population limit. The obtained timescales were: 646 ± 13 fs for ODM2/MR-CISD/LZSH trajectories, 705 ± 22 fs for AIQM1/MR-CISD/LZSH, and 515 ± 9 fs for ODM2/MR-CISD/FSSH simulations. This quantifies the agreement between the methods used and shows that the population dynamics is more sensitive to the electronic-structure model used for single-point calculations rather than the surface hopping algorithm, similar to the previously reported analysis.[13]

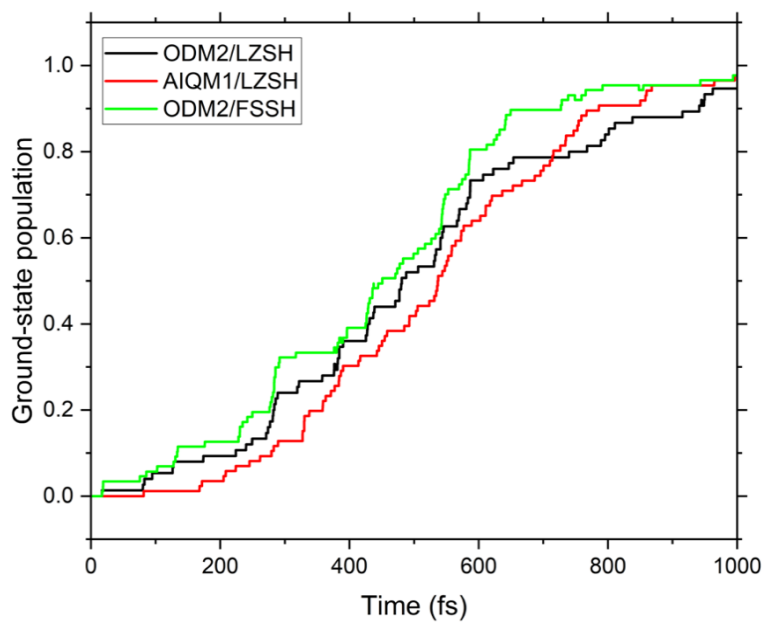

**Figure 9.** Evolution of the ground state population during the photoisomerization dynamics of *trans*-azobenzene. The black curve represents results from ODM2/LZSH simulations, the red curve is the population from AIQM1/LZSH dynamics, and the green curve is obtained from ODM2/FSSH simulations.

As a next step in this investigation, one can look at the predicted quantum yields of *trans* → *cis* photoisomerization extracted from each set of simulations. The simulated quantum yield Φ is calculated as the fraction of the trajectories relaxing to form *cis* azobenzene $N_{cis}$, to the total number of trajectories $N_{traj}$: $\Phi = N_{cis}/N_{traj}$. As a descriptor of the formed photoproduct, dihedral angle $C_1$–$N_2$–$N_3$–$C_4$ (Ψ) of azobenzene was used (as defined in Figure 8), which has the value of approximately 180 degrees for *trans*-azobenzene, and close to 0 degrees for *cis*-azobenzene.

The distribution of the Ψ angle at the trajectories' final point can be found in Figure 10 for all 3 sets of trajectories. Taking Ψ < 50° as an indicator of a successful *trans* → *cis* photoisomerization, we can calculate the quantum yields Φ as 0.09 ± 0.07 for ODM2/LZSH simulations, 0.24 ± 0.09 for AIQM1/LZSH and 0.08 ± 0.06 for ODM2/FSSH. The uncertainties of the quantum yields were estimated using the normal approximation interval for a binomial process, assuming a confidence interval of 95%. Analyzing those results, one can notice that both sets of trajectories, where the electronic structure was computed at the ODM2 level of theory, predict a similar quantum yield of 0.08 – 0.09. At the same time, AIQM1 gives a different result of 0.24. This further exemplifies how the electronic structure model used for single-point calculations can profoundly impact the results. Remarkably, it is the AIQM1 simulations that predict the correct quantum yield of this photoprocess, as experimental results typically report a value of 0.2 to 0.36,[95, 96] after an $n \to \pi^*$ excitation. This shows the potential of AIQM1 to deliver high-quality results for excited-state calculations, as well as the possibility of obtaining accurate results using LZSH simulations.

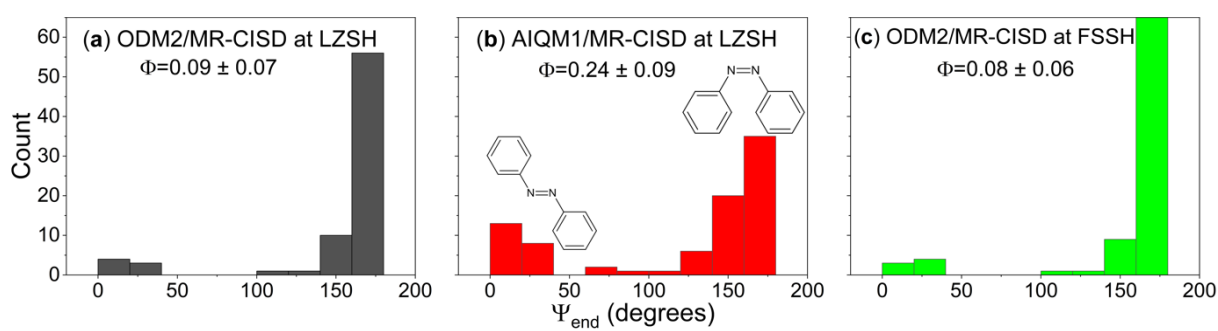

**Figure 10.** Distribution of the C–N–N–C dihedral angle of azobenzene at the trajectories final points: (a) ODM2/LZSH; (b) AIQM1/LZSH; (c) ODM2/FSSH.

Lastly, we will discuss the numerical stability and cost of the simulations. The ODM2/FSSH calculations performed using the MNDO99 program took approximately 2 CPU hours per trajectory, with 87 out of the 100 trajectories successfully propagating for 1000 fs. For ODM2/LZSH, the cost was the same as for FSSH, with 75 out of 100 trajectories propagating until the assumed simulation end time. Finally, the AIQM1/LZSH calculations had a higher computational cost of about 8 CPU hours per trajectory, with 86 trajectories from that set completed without errors. All calculations were performed using Intel Xeon Platinum 8268 CPUs. It is also important to note that the LZSH trajectories ran in MLatom were propagated in parallel on a single multi-threaded CPU node and did not require submission to multiple nodes.

*Molecular ferro-wire*

To demonstrate the capability of LZSH dynamics to describe large photoactive systems, we have performed NAMD simulations for an electrically polarized molecular wire, which has been studied before with the semi-empirical QM method and fewest-switches surface hopping dynamics[97] (Figure 11). This system is fascinating due to its ability to perform spontaneous charge carrier separation upon photoexcitation. However, its size — 80 atoms, including 54 heavy ones — makes it computationally tricky and expensive to study with NAMD methods, even at the semi-empirical level of theory.

To investigate the performance of LZSH for describing the excited state dynamics of such system, four sets of trajectories were propagated: (a) LZSH with ODM2/CIS electronic structure treatment; (b) LZSH with AIQM1/CIS; (c) fewest-switches surface hopping (FSSH) with ODM2/CIS (d) FSSH with ODM2/MR-CISD. Fifty trajectories were propagated in each set for 200 fs with a 0.5 fs time step, starting from the same initial conditions sampled and filtered from canonical ground-state ODM2 dynamics, as described in the previous study.[97] Detailed settings of the methods used can be found in the Computational Details section.

The resulting curves, illustrating the electronic state population evolution during the dynamics, can be found in Figure 11. Generally, the results obtained with all tested approaches stay in agreement. More specifically, in all the LZSH simulations, none of the trajectories relax to the ground state, which is the system's correct, expected physical behavior. To compare those results qualitatively, we again use an exponential function fit$_1$ population rise, with $f(t)$ defined as for the azobenzene dynamics. The resulting timescales were 73.2 ± 1.5 fs for LZSH at ODM2/CIS, 40.26 ± 0.76 fs for LZSH at AIQM1/CIS, 41.92 ± 0.56 fs for FSSH at ODM2/CIS and 36.15 ± 0.35 fs for FSSH at ODM2/MR-CISD. Interestingly, the AIQM1/CIS method for electronic structure treatment provided a timescale closest to the semi-empirical MRCI calculations performed before.[97]

Lastly, one can compare the computational costs of these methods on Intel Xeon Platinum 8268 CPUs. For the FSSH simulations with ODM2/CIS, the cost of propagating all 50 trajectories was about 38 CPU hours. However, with the more accurate ODM2/MR-CISD, the cost grew to about 52 CPU-hours per trajectory, for a total of almost 2600 CPU-hours. Propagating the LZSH dynamics at the ODM2/CIS level of theory took about 67 CPU hours and 107 CPU hours for AIQM1/CIS. It should be noted that our implementation of LZSH can propagate all the trajectories in parallel, making the calculations particularly convenient.

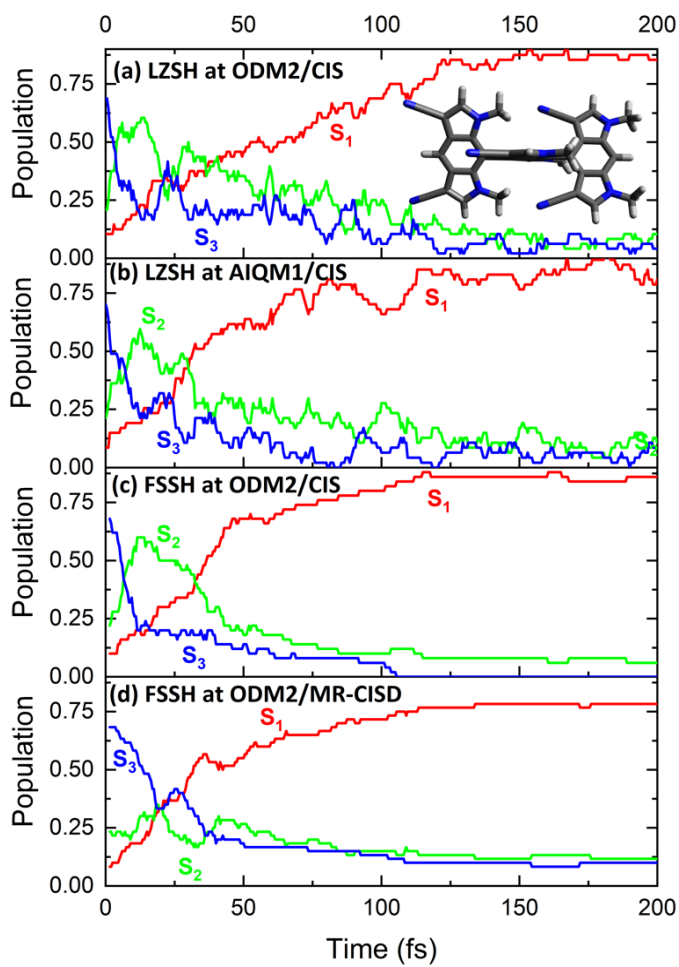

**Figure 11.** Electronic state population evolution in the nonadiabatic molecular dynamics of the polarized molecular wire (structure shown as inset in top right), obtained using different computational schemes: a) LZSH at ODM2/CIS, b) LZSH at AIQM1/CIS, c) FSSH at ODM2/CIS, d) FSSH at ODM2/MR-CISD. The $S_1$, $S_2$, and $S_3$ states are marked in red, green, and blue, respectively.

**Conclusions and outlook**

In this article, we introduced surface-hopping molecular dynamics based on the MLatom ecosystem. It is based on the versatile Python API that enables the construction of the entire workflow for surface-hopping simulations, starting from the initial molecular structure to initial conditions sampling and ending with the population plots. The simulations currently can be performed using the LZBL algorithm following the Landau–Zener approximation based on the Belyaev–Lebedev reformulation for the adiabatic representation. The initial conditions can be sampled from the Wigner distribution using the routines adapted from Newton-X.

The ecosystem supports various quantum-mechanical and machine learning approaches, including their combinations, such as the universal ML-improved QM method AIQM1. Our results show that AIQM1 is an attractive method for cost-effective surface-hopping simulations, and it does not need any ML training or active learning, further simplifying its use. AIQM1 provides quantum yields in a better agreement for azobenzene photoisomerization than its related semi-empirical QM methods at a similar cost.

All implementations in MLatom described here are available open source, with detailed tutorials and scripts that can be easily adjusted for the requested type of simulations. Some of the calculations (currently, with ML models) can be also performed via a web browser on the XACS cloud computing platform.[98]

The ecosystem's versatility enabled us to develop protocols for active learning, which will be reported separately. In addition, we are working on a deeper and more efficient integration of Newton-X's vast range of functionalities for surface hopping dynamics, such as fewest-switches surface hopping, to facilitate similar workflows via the Python API.

**Computational details**

*Nonadiabatic dynamics of methaniminium cation*

The initial conditions were generated at the B3LYP/def2-SVPP level from the unfiltered Wigner sampling. In the case of methaniminium cation, the active space of the SA-3-CASSCF(12,8) method is composed of 8 active orbitals (6 occupied and 2 unoccupied molecular orbitals) and 12 active electrons. The MCSCF calculations were performed using Columbus interfaced with MLatom. The AIQM1/MR-CISD(4,3) calculations were done using MNDO interfaced with MLatom. General configuration interaction based on the graphical unitary group approach (GUGA-CI) is used for MR-CISD(4,3). The total number of occupied orbitals and unoccupied orbitals in the active space is set to be 2 and 1, respectively. The MNDO option `movo` is set to −1 to automatically include π orbitals in the active space.

*Pyrazine*

The initial conditions were generated at the MP2/cc-pVDZ level from the Wigner sampling filtered using the excitation energy window at ca. 5.2 eV (at 300 K) with Newton-X. For ML dynamics, we trained the individual ANI models for each electronic state on both energies and gradients using the 90:10 splitting for the sub-training and validation sets and other default settings in MLatom. The models were trained on the unfiltered 250 initial conditions.

*Isomerization dynamics of trans-azobenzene*

All ODM2/MR-CISD calculations for azobenzene, including the semi-empirical part of the AIQM1 method, used the half-electron restricted open-shell Hartree–Fock formalism[99] in the initial SCF step, with the HOMO and LUMO orbitals singly occupied. This configuration was then supplemented with two additional references: a closed-shell HOMO-HOMO configuration and a doubly-excited LUMO-LUMO configuration. The active space comprised 8 electrons in 10 orbitals (four highest occupied and six lowest unoccupied). The MRCI procedure accounted for single and double excitations within a such-defined active space.

The fewest-switch surface hopping calculations were benchmarked using the MNDO99 program.[72] All relevant properties, such as the nonadiabatic coupling vectors, were computed *on the fly*. Nuclear motion was propagated with a time step of 0.1 fs. In comparison, electronic structure integration was performed with a time step of 0.0005 fs. Decoherence was accounted for using the simplified decay of mixing decoherence correction,[100] with a standard value of 0.1 au. Dynamics was propagated for 1000 fs. After hopping, the nuclear velocities were rescaled in the direction of the nonadiabatic coupling vector.

LZSH dynamics were performed using the same settings as FSSH, with the velocities rescaled in the momentum direction using the reduced kinetic energy reservoir described in the Methods section.

We set the frequencies smaller than 100 cm$^{-1}$ to 100 cm$^{-1}$ to generate initial conditions via the Wigner sampling as described in the Methods section.

*Nonadiabatic dynamics of a polarized molecular ferro-wire*

All NAMD simulations were performed using the same settings outlined for azobenzene, except for the time step, which was set to 0.5 fs for nuclear motion and 0.0025 fs for electronic structure integration. In LZSH dynamics, the velocity adjustment was performed using the full kinetic energy reservoir due to the large and modular nature of the simulated system.

**Data availability**

Trajectories are available as described at https://github.com/dralgroup/lzsh24.

**Code availability**

The code is available as open source under the MIT license at https://github.com/dralgroup/mlatom. The described calculations can be performed with the ≥3.3.0 version of the MLatom, and the corresponding open-source scripts and tutorials under the MIT license are available at https://github.com/dralgroup/lzsh24.

**Author contributions**

All authors shaped this research project. L.Z. implemented the LZBL surface hopping dynamics and window selection of the initial conditions with input from P.O.D., S.V.P., and M.B. She also implemented the calculation of excited-state properties via interfaces to COLUMBUS and MNDO (the latter one with the help from F.G.). S.V.P. implemented the initial version of the Turbomole interface for the ADC(2) calculations, later modified by L.Z. M.M. performed simulations and analysis for azobenzene and polarized molecular wire under the supervision of J.J. and P.O.D. F.G. implemented the electronic state information in the data formats co-designed with P.O.D. Y.-F.H. implemented the initial conditions sampling from the Wigner distribution by adapting Newton-X code to the MLatom format. P.O.D. also suggested the treatment for sampling from low-frequency normal models, which was tested by Y.H. and M.M. Y.C. implemented frequency calculations via the interface to PySCF. L.Z., M.M., J.J., and P.O.D. co-wrote the original version of the manuscript with input from other authors. F.G. and Y.-F.H. also contributed to the preparation of the figures.

**Acknowledgments**

L.Z. thanks Peikun Zheng for technical help with MNDO.

P.O.D. acknowledges funding by the National Natural Science Foundation of China (No. 22003051 and funding via the Outstanding Youth Scholars (Overseas, 2021) project), the Fundamental Research Funds for the Central Universities (No. 20720210092). This project is supported by Science and Technology Projects of Innovation Laboratory for Sciences and Technologies of Energy Materials of Fujian Province (IKKEM) (No: RD2022070103).

M.M. and J.J. would like to acknowledge the Polish Ministry of Education and Science for funding this research under the program "Perły Nauki," grant number PN/01/0064/2022, amount of funding and the total value of the project: 239 800,00 PLN, as well as the Polish high-performance computing infrastructure, PLGrid (HPC Centers: ACK Cyfronet AGH) for providing computer facilities and support within computational grant no. PLG/2023/016750.

S.V.P. and L.C. acknowledge support from the Key Research Project of Zhejiang Lab (No. 2021PE0AC02).

M.B. thanks the European Research Council (ERC) Advanced grant SubNano (grant agreement 832237).